\documentclass[twocolumn,showpacs,preprintnumbers,amsmath,amssymb]{revtex4}

\usepackage{graphicx}
\usepackage{dcolumn}
\usepackage{bm}
\usepackage{color}

\begin{document}


\title{Spin Model of O$_2$-based Magnet 
in a Nanoporous Metal Complex}

\author{M. Soda$^{1}$, Y. Honma$^{1}$, S. Takamizawa$^{2}$, 
S. Ohira-Kawamura$^{3}$, K. Nakajima$^{3}$, and T. Masuda$^{1}$}
\affiliation{%
$^{1}$Neutron Science Laboratory, Institute for Solid State Physics, 
University of Tokyo, Tokai, Ibaraki 319-1106, Japan\\
$^{2}$Department of Nanosystem Science, Graduate School of 
Nanobioscience, Yokohama City University, Kanazawa-Ku, Yokohama, 
Kanagawa 236-0027, Japan\\
$^{3}$Materials and Life Science Division, J-PARC Center, Tokai, 
Ibaraki 319-1195, Japan
}%

\date{\today}

\begin{abstract}
Inelastic neutron scattering experiments are performed 
on a nanoporous metal complex 
Cu-Trans-1,4-Cyclohexanedicarboxylic Acid (Cu-CHD) 
adsorbing O$_2$ molecules to identify the spin model 
of the O$_2$-based magnet realized in the host complex. 
It is found that 
the magnetic excitations of Cu-CHDs adsorbing low- and high-concentration 
O$_2$ molecules are explained by different spin models, 
the former by spin dimers and the latter by spin trimers. 
By using the obtained parameters and also by assuming that 
the levels of the higher energy states 
are reduced because of the non-negligible 
spin dependence of the molecular potential of oxygen, 
the magnetization curves are explained in 
quantitative level. 
%
\end{abstract}

\maketitle
\section{Introduction}
Natural oxygen, the second abundant constituent in the air, 
is a magnet having spin $S$=l induced by a couple of 
${\pi}_g$ electrons. 
The oxygen molecules are condensed and crystalized at 54 K 
by van-der Waals interaction. 
The solid oxygen realized at the low temperatures include 
${\alpha}$ phase with monoclinic $C2/m$ at 
$T \le 23$ K, ${\beta}$ phase 
with hexagonal $R\bar{3}m$ at $23~{\rm K} < T \le 43$ K, 
and ${\gamma}$ phase with cubic $Pm3n$ at 
$43~{\rm K} < T \le 54$ K.\cite{Uyeda,Freimana} 
Since the magnetic system is phenomenologically coupled to the lattice 
system thorough magneto-elastic coupling, the successive phase transition is 
magneto-structural transition accompanied by 
the change of the spin Hamiltonian. 
Magnetic susceptibility measurements suggested that 
quasi-two-dimensional spin system in the $\alpha$ phase, 
triangular spin system in the $\beta$ phase, 
and one-dimensional spin system in the $\gamma$ phase 
were realized.\cite{Uyeda} 
Applying pressure induces more phases including metal,\cite{Desgreniers} 
superconductivity,\cite{Shimizu}, and an insulating 
magnetically ordered phases.\cite{Goncharenko} 
The variety of the phases suggests that the energy scales of the 
coupling among multi-degrees of freedom, {\it i.e.,} 
lattice, spin, and orbital, 
are close to each other in the soft solid crystalized by 
Van der Waals interaction. 
Indeed the molecular potential between a pair of O$_2$ molecules is strongly 
dependent on the spin state, which means that the geometrical 
configuration of the O$_2$ molecules and the spin state 
are closely correlated.\cite{Hemert,Wormer,Bussery,Bussery94,Nozawa,Obata} 
To manipulate the oxygen molecule O$_2$ and to artificially synthesize 
a novel type of O$_2$-based magnet is, therefore, the challenge 
in the new field of magnetism. 

Pioneering work is found in the adsorbed 
O$_2$ on the surface of graphite, where triangular lattice 
of O$_2$ is realized.\cite{McTague,Murakami} 
Combination of the magnetic susceptibility and neutron diffraction 
measurements 
reveals the magnetically ordered state at low temperatures. 
Another root for the O$_2$-based magnet is to utilize nano-materials 
such as microporous metal 
complexes,\cite{Kitagawa,Takamizawa,Takamizawa2,Kondo,Mori,Masuda}
nanoporous silica,\cite{Wallacher,AckermannEPL} 
or carbon nanotubes.\cite{Hagiwara} 
The adsorbed molecules staying at the minimum of the Van der Waals 
potential in the nanopore form a supercrystal, 
leading to the realization of the O$_2$-based magnet. 

We focus our attention on a metal complex having 
one-dimensional nanopores, Cu-Trans-1,4-Cyclohexanedicarboxylic 
Acid Cu$_2$(OOC-C$_6$H$_4$-COO) abbreviated as Cu-CHD.\cite{Mori} 
The adsorbed O$_2$ molecules form dimer-like structure 
in the nanopores.\cite{Hori} 
Indeed, the magnetic susceptibility 
of Cu-CHD adsorbing 0.18 mole of O$_{2}$ per half of formula unit showed 
a rapid decrease with the decrease of the temperature, 
which is consistent with a spin dimer model. 
It was, however, explained not by $S$=1 spin dimer 
but rather by $S$ = 1/2 spin dimer. 
To explain the unusual bulk property, 
the spin-dependent 
Van der Waals potential\cite{Hemert,Wormer,Bussery,Bussery94}
 was discussed, 
and the anomalous energy spectrum 
was proposed.\cite{Hori} 
The precise energy scheme is to be directly revealed by a spectroscopic 
technique. 

The magnetic susceptibility of Cu-CHD adsorbing 
high concentration of O$_{2}$ molecules 
showed qualitatively different behavior from that of 
the low concentration one. 
Curie-Weiss like behavior was enhanced in 
the low $T$ region.\cite{Mori} 
The result suggested that the spin model and 
the corresponding energy scheme can be tuned 
by the concentration of 
the O$_2$ molecules in the O$_2$-based magnet. 
The detailed model is to be experimentally identified. 

In the present study 
we carried out inelastic neutron scattering (INS) experiments 
at low temperatures 
in order to observe the low-energy excitations and to 
clarify the spin model of the O$_2$-based magnet 
realized in Cu-CHD. 
We found that the 
magnetic excitation of Cu-CHD adsorbing less 
O$_{2}$ is explained by a spin-dimers model of which 
the exchange constants are normally distributed. 
In contrast, that of Cu-CHD adsorbing more O$_{2}$ 
is explained by a spin-trimers model with the distributed 
exchange constants. 
It was found that the spin model is tuned by the 
concentration of O$_2$ in this system. 
Based on the spin models with the conclusive parameters 
determined by INS experiments and by introducing the 
reduction of the energy levels of higher states, 
we quantitatively explained the 
magnetization curves. 
The effect of the spin-dependent 
Van der Waals potential to the anomalous energy scheme 
is experimentally confirmed. 

\section{Experimental details}

Polycrystalline Cu-CHD was synthesized in methanol solution. The details 
are described elsewhere.\cite{Inoue} 
The total mass was 5.93 g. 
Since Cu-CHD includes many hydrogen atoms that have 
large incoherent neutron scattering cross sections, we paid 
special attention to extract the net contribution of the 
adsorbed oxygen molecules. 
The first step was to prepare the bare Cu-CHD sample in which 
any type of guest molecules including H$_2$O, N$_2$, O$_2$, etc., 
were eliminated. 
The second step was to measure the INS spectrum of the bare Cu-CHD sample 
as background data. 
The third step was to prepare the oxygen adsorbed Cu-CHD sample. 
Here we used exactly the same 
sample that was used for the background measurement in the second step. 
The forth step was to measure the INS spectrum of the 
oxygen adsorbed Cu-CHD. 
Ideally the net contribution of the adsorbed oxygen 
can be extracted by subtracting the INS spectrum in the second step 
from that in the forth step. 
In reality we found additional intensity that increased with 
the increase of the wave number transfer $Q$ even after the 
background subtraction process. 
In addition we found flat intensity that was independent both 
on $Q$ and the energy transfer $\hbar \omega$. 
During the analyses on the data after background subtraction, 
therefore, we used {\it additional background} 
of which the form was $A({\hbar}{\omega})Q^2 + B$. 
Here $A({\hbar}{\omega})$ and $B$ were free parameters. 
We could not identify the origin of the additional background, 
even though we could make some speculations; 
$A({\hbar}{\omega})$ term 
would be from the deformation of the host compound 
after the oxygen adsorption, 
and $B$ term would be from 
the difference of white background such as electronic noise 
between before and after the oxygen adsorption. 

The procedure of preparing the bare Cu-CHD is as follows.
The Cu-CHD sample in the Aluminum made container 
was warmed at about 105 $^{\circ}$C and was evacuated by 
a vacuum pump with monitoring the pressure. 
We used the gas evacuation/introduction system described in 
Fig. 2 in Ref.\onlinecite{Masuda}
We waited until the pressure reached a few Pa, which is the 
capacity limit of the vacuum pump, and we confirmed that 
the Cu-CHD sample had discharged all the molecules. 
We, then, seal the container by closing a V1 valve in Fig. 2 
in Ref.\onlinecite{Masuda}, put the sample in a liquid helium cryostat, 
cool the temperature down to 4.5 K, and collected INS spectrum 
of the bare Cu-CHD. 
After the background measurement 
we set the temperature of the bare Cu-CHD at 110 K 
and introduced O$_{2}$ gas in the nanopore by using a buffer container 
of which the capacities were known. 
The pressure of the gas was kept below 8$\times$10$^{4}$ Pa.
The volume of the introduced oxygen was estimated by 
the capacity of the buffer container and the pressure of the 
O$_2$ gas inside the container. 
After the O$_2$ introduction procedure 
the sample container was sealed, and was cooled down to 4.5 K.
All measurements were carried out at $T$=4.5 K.
We prepared two samples 
that adsorbed different amount of O$_2$ 
in order to examine the O$_2$ concentration dependence 
of the spin model realized in the nanopores. 
Exactly the same polycrystalline Cu-CHD was 
used as the host compound for both samples. 
In prior to the preparation of each sample 
we discharged the gas in the nanopore 
at 105 $^{\circ}$C according to the procedure described above. 
The O$_{2}$ mole ratios sealed 
in the sample container were O$_{2}$/(Cu atom)=0.3${\pm}0.07$ 
and 2.0${\pm}0.4$, which are labeled as 0.3O$_{2}$-(Cu-CHD) 
and 2.0O$_{2}$-(Cu-CHD), respectively.

Neutron scattering experiment was carried out using 
the cold neutron chopper spectrometer AMATERAS installed 
at J-PARC.\cite{Nakajima,Inamura} 
The initial neutron energies $E_{i}$=3.135 meV
 and $E_{i}$=7.743 meV were used,
 and their energy resolutions are about 0.15 and 0.45 meV 
at the elastic position. 

\section{Results}
\begin{figure}
\begin{center}\leavevmode
\includegraphics[width=7 cm]{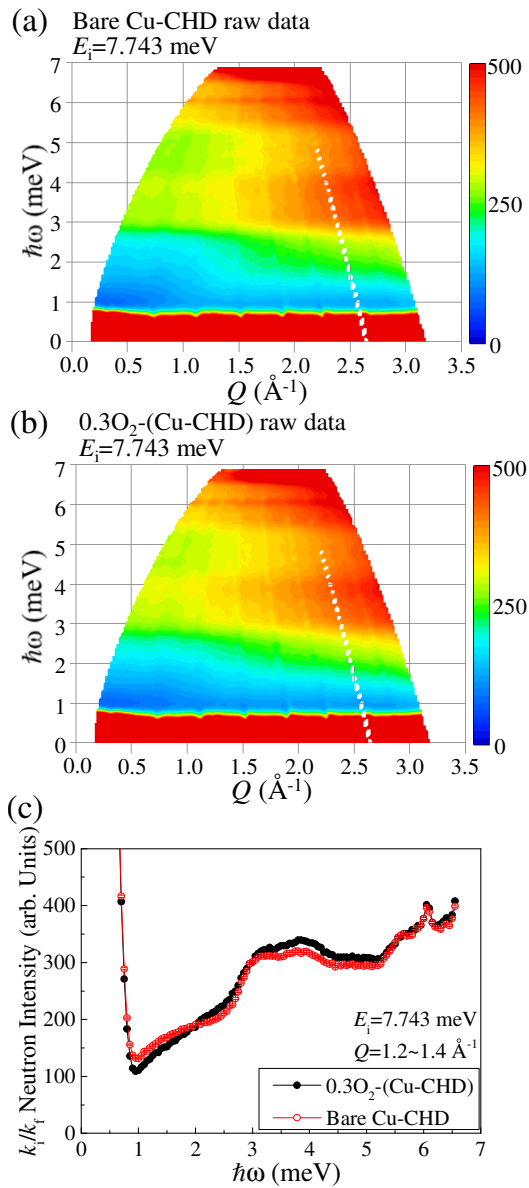}
\caption{
(Color online)
(a) Contour map of the raw neutron spectrum bare-(Cu-CHD)
 by using $E_{i}$=7.743 meV. 
(b) Contour map of the raw neutron spectrum 0.3O$_{2}$-(Cu-CHD)
 by using $E_{i}$=7.743 meV. 
(c) One-dimensional energy cut obtained by integrating the intensity 
in the range of $1.2~{\rm \AA}^{-1} \le Q  \le 1.4~{\rm \AA}^{-1}$ 
for raw data of 0.3O$_{2}$-(Cu-CHD) and bare Cu-CHD. 
}
\label{Fig. 1}
\end{center}
\end{figure}

\begin{figure*}
\begin{center}\leavevmode
\includegraphics[width=14 cm]{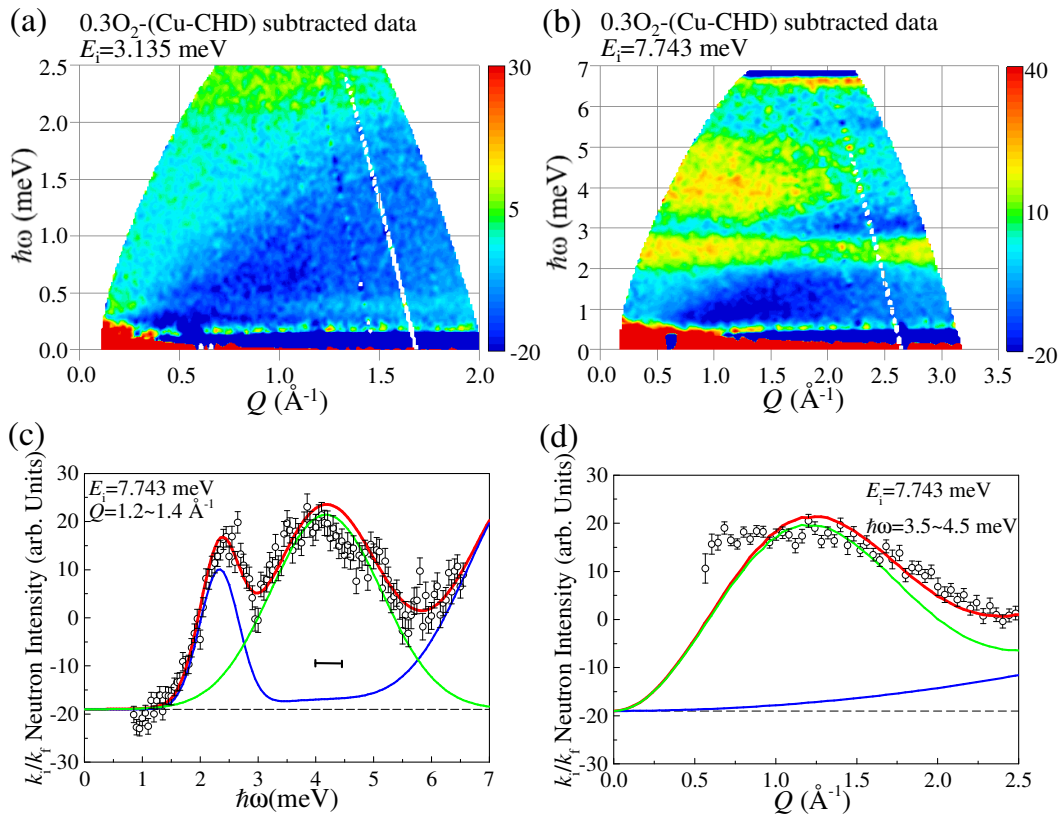}
\caption{
(Color online)
(a,b) Contour map of the neutron spectrum
of 0.3O$_{2}$-(Cu-CHD) after the background subtraction 
by using (a) $E_{i}$=3.135 meV and (b) $E_{i}$=7.743 meV. 
(c,d) Neutron intensity for 0.3O$_{2}$-(Cu-CHD) 
after the background subtraction by using $E_{i}$=7.743 meV. 
The green curve is the neutron cross section of a group of 
spin dimers having distributed $J$. 
The blue curve is the additional background 
having $Q$ dependence.  
The red curve is the sum of green and blue curves. 
The dashed line shows the additional flat background. 
(c) One-dimensional energy cut obtained by integrating intensity 
in the range of $1.2~{\rm \AA}^{-1} \le Q  \le 1.4~{\rm \AA}^{-1}$. 
The energy resolution is shown by the horizontal bar. 
(d) One-dimensional $Q$ cut obtained by integrating intensity 
in the range of $3.5~{\rm meV} \le E  \le 4.5~{\rm meV}$.
}
\label{Fig. 1a}
\end{center}
\end{figure*}

\begin{figure*}
\begin{center}\leavevmode
\includegraphics[width=18 cm]{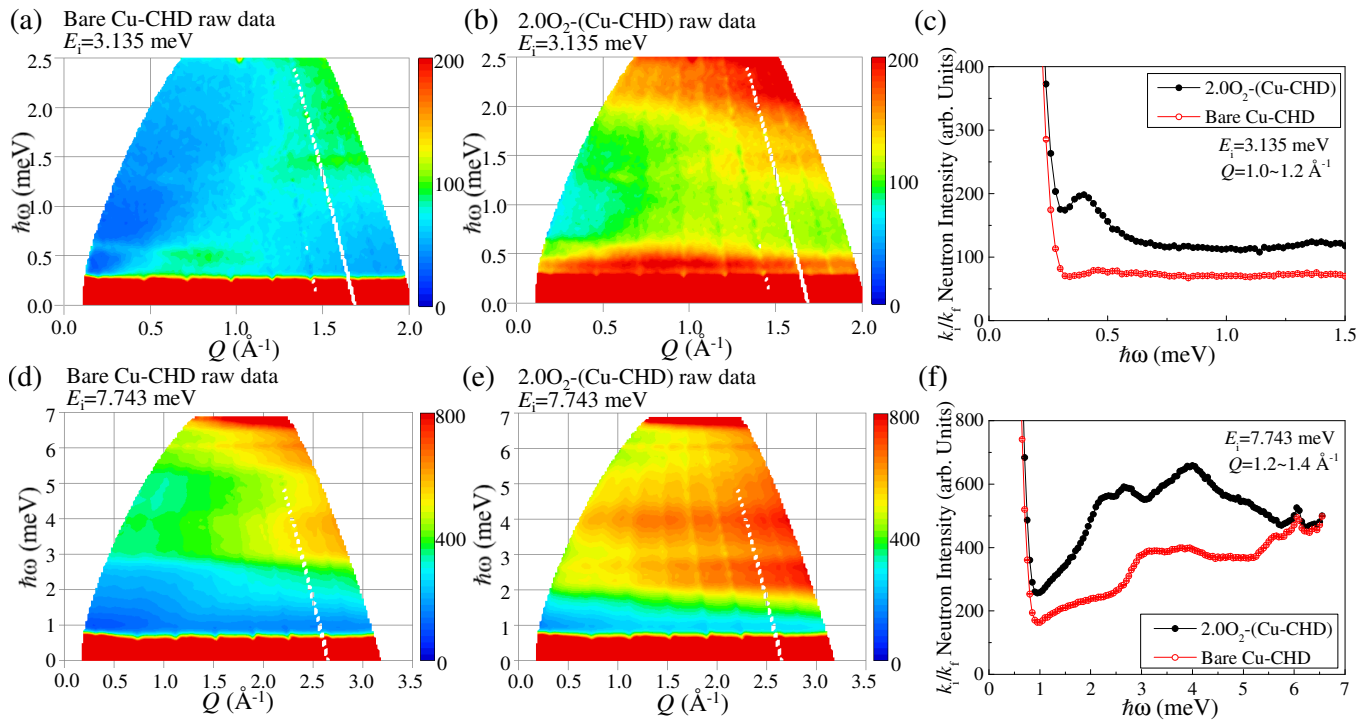}
\caption{
(color online)
(a,b) Contour map of the raw neutron spectrum
of (a) bare Cu-CHD and (b) 2.0O$_{2}$-(Cu-CHD) 
by using $E_{i}$=3.135 meV. 
(c) One-dimensional energy cut in the low-energy region obtained 
by integrating intensity 
in the range of $1.0~{\rm \AA}^{-1} \le Q  \le 1.2~{\rm \AA}^{-1}$ 
for raw 2.0O$_{2}$-(Cu-CHD) and bare Cu-CHD by 
using $E_{i}$=3.135 meV. 
(d,e) Contour map of the raw neutron spectrum
of (d) bare Cu-CHD and (e) 2.0O$_{2}$-(Cu-CHD) 
by using $E_{i}$=7.743 meV. 
(f) One-dimensional energy cut in the high-energy region obtained 
by integrating intensity 
in the range of $1.2~{\rm \AA}^{-1} \le Q  \le 1.4~{\rm \AA}^{-1}$ 
for raw 2.0O$_{2}$-(Cu-CHD) and bare Cu-CHD by using 
$E_{i}$=7.743 meV. 
}
\label{Fig. 2}
\end{center}
\end{figure*}

\begin{figure*}
\begin{center}\leavevmode
\includegraphics[width=18 cm]{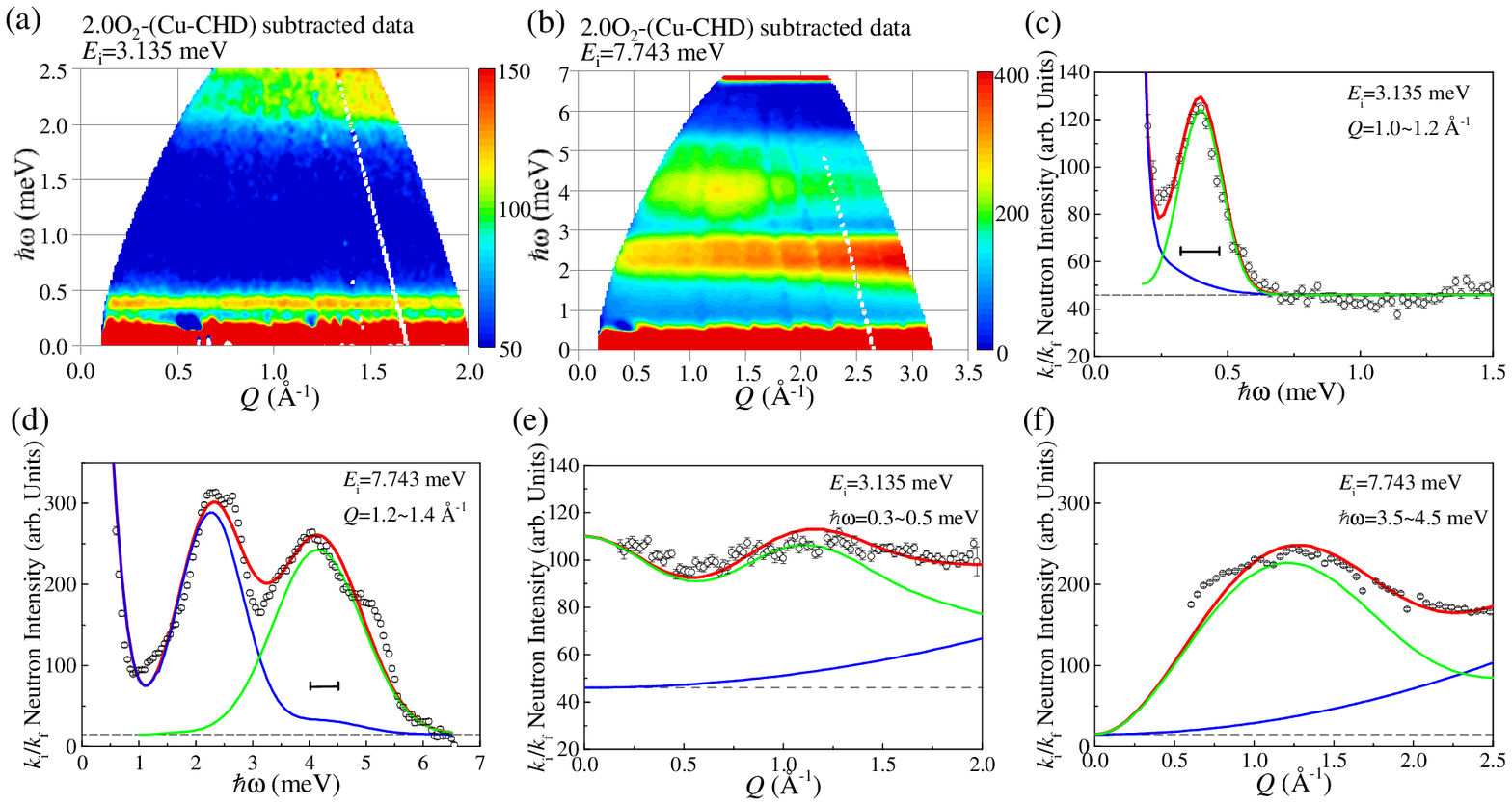}
\caption{
(color online)
Neutron intensity for 2.0O$_{2}$-(Cu-CHD) 
after the background subtraction. 
The green curve is the magnetic excitation, 
the blue curve is the additional background 
having $Q$ dependence, 
and the red curve is the sum of green and blue curves. 
The dashed line shows the background. 
(a) One-dimensional energy cut in the low-energy region 
obtained by integrating intensity 
in the range of $1.0~{\rm \AA}^{-1} \le Q  \le 1.2~{\rm \AA}^{-1}$. 
The energy resolution is shown by the horizontal bar. 
(b) One-dimensional energy cut in the high-energy region 
obtained by integrating intensity 
in the range of $1.2~{\rm \AA}^{-1} \le Q  \le 1.4~{\rm \AA}^{-1}$. 
(c) One-dimensional $Q$ cut obtained by integrating intensity 
in the range of $0.3~{\rm meV} \le E  \le 0.5~{\rm meV}$. 
The energy resolution is shown by the horizontal bar. 
(d) One-dimensional $Q$ cut obtained by integrating intensity 
in the range of $3.5~{\rm meV} \le E  \le 4.5~{\rm meV}$.
}
\label{Fig. 2a}
\end{center}
\end{figure*}

Figures \ref{Fig. 1}(a) and (b) show the raw inelastic 
neutron scattering 
(INS) spectra of bare Cu-CHD and 
 0.3O$_{2}$-(Cu-CHD), respectively. 
The initial energy $E_{i}$ is 7.743 meV. 
Strong incoherent scattering from hydrogen atoms and/or 
functional groups 
were observed
 in the wide region in the energy transfer ($\hbar \omega$) - 
wave number transfer ($Q$)
space in both panels. 
Figure \ref{Fig. 1}(c) shows typical one-dimensional (1D) energy cut 
obtained
 by integrating the intensity in the range of 
$1.2~{\rm \AA}^{-1} \le Q  \le 1.4~{\rm \AA}^{-1}$
 for the raw spectrum of 0.3O$_{2}$-(Cu-CHD) and that of the bare Cu-CHD.
Statistically meaningful increases are observed
 at about 2.5 meV and 4 meV in 0.3O$_{2}$-(Cu-CHD). 
The enhancement of the intensity is due to 
 the additional scattering of the adsorbed oxygen molecules. 
The decrease of the intensity
is observed at about 1 meV in Fig. \ref{Fig. 1}(b).
As we described in the Section II the INS spectrum of 
the bare Cu-CHD is regarded as 
background, but its intensity is comparable to 
the 0.3O$_{2}$-(Cu-CHD) because of the large incoherent scattering 
from the hydrogen atoms included. 
Even though the careful background measurement, 
the background has rather larger intensity 
than 0.3O$_{2}$-(Cu-CHD), of which the 
origin would be unavoidable artifacts. 


Figure \ref{Fig. 1a}(a) shows the INS spectrum of 0.3O$_{2}$-(Cu-CHD)
 after the background subtraction in the low-energy
 region for the data set of $E_{i}$=3.135 meV. 
The absence of excitation at $\hbar \omega \lesssim 2$ meV is confirmed in the 
experimental error. 
Figure \ref{Fig. 1a}(b) shows the INS spectrum of 0.3O$_{2}$-(Cu-CHD) 
in the high energy 
region for the data set of $E_{i}$=7.743 meV 
after the background subtraction. 
The dispersionless excitations are observed
 at $\hbar \omega$$\simeq$2.4 meV and 4 meV.
These two excitations are also probed as two peaks 
in the 1D-energy cut 
after the background subtraction shown in Fig. \ref{Fig. 1a}(c).

To examine the $Q$-dependence of the excitation at 
$\hbar \omega$$\simeq$4 meV, 
the 1D-$Q$ cut obtained by integrating 
the intensity in the range of 
$3.5~{\rm meV} \le \hbar \omega \le 4.5~{\rm meV}$ 
is shown in Fig. \ref{Fig. 1a}(d). 
The intensity of the excitation has a kink at about 1.2 \AA$^{-1}$ 
and it decreases with increasing $Q$. 
The dispersionless behavior means that the origin of the 
excitation is a cluster. 
The suppressed intensity at large $Q$ means that 
the excitation is dominated by a magnetic scattering. 
The excitation induced by the oxygen adsorption is, 
therefore, the magnetic one from some types of the O$_2$ molecules cluster. 
The kink at finite $Q$ means that the spin correlation is antiferromagnetic, 
and the maximum $Q$ exhibits the inversed length-scale of the intra-cluster 
distance. 
As for the excitation at $\hbar \omega$$\simeq$2.4 meV 
in Fig. \ref{Fig. 1a}(b), 
the intensity shows monotonic decrease with the increase of 
$Q$ in the range of $Q \lesssim 2.2 {\rm \AA}^{-1}$ 
and, then, it increases in the range of $2.2 {\rm \AA}^{-1} \lesssim Q$. 
We considered that this excitation  
is not intrinsically magnetic one, which will be discussed later.

Raw INS spectra of the bare Cu-CHD and the 2.0O$_2$-(Cu-CHD) 
with $E_i = 3.135$ meV are shown in Fig.~\ref{Fig. 2} (a) and (b), 
respectively. 
A flat excitation at $\hbar \omega \sim 0.4$ meV is newly observed 
in the 2.0O$_2$-(Cu-CHD) in addition to smeared intensities in 
wide $\hbar \omega$ - $Q$ space. 
1D-energy cuts of the 2.0O$_{2}$-(Cu-CHD) and the bare Cu-CHD 
in the integration range of 1.0 \AA $^{-1} \le Q \le $ 1.2 \AA $^{-1}$ 
is shown in Fig.~\ref{Fig. 2} (c). 
Peak structure at 0.4 meV is clearly observed. 
Raw INS spectra of the bare Cu-CHD and the 2.0O$_2$-(Cu-CHD) 
with $E_i = 7.743$ meV are shown in Fig.~\ref{Fig. 2}(d) and (e), 
respectively. 
The spectrum in Fig.~\ref{Fig. 2}(d) is the same data as 
shown in Fig.~\ref{Fig. 1}(a) but the intensity scale is different. 
Broad excitations at 3 meV $\lesssim \hbar \omega \lesssim$ 5 meV 
in low $Q$ region are newly observed in 2.0O$_2$-(Cu-CHD). 
The 1D-energy cuts 
for $E_i$ = 7.743 meV is shown in Fig.~\ref{Fig. 2} (f). 
The Cu-CHD adsorbing more O$_{2}$ molecules exhibits 
the large enhancement of the neutron intensity 
compared to those of 0.3O$_{2}$-(Cu-CHD). 

The INS spectrum of 2.0O$_{2}$-(Cu-CHD) by using the $E_{i}$=3.135 meV 
after the background subtraction is shown in Fig. \ref{Fig. 2a}(a). 
The dispersionless excitation is observed at $E$$\simeq$0.4 meV.
The 1D-energy cut in Fig. \ref{Fig. 2a}(c) 
exhibits that the energy width of the excitation 
is almost the resolution limited.
The INS spectra of 2.0O$_{2}$-(Cu-CHD) by using the $E_{i}$=7.743 meV 
after the background subtraction is shown in Fig. \ref{Fig. 2a}(b). 
The dispersionless excitations are observed at 
$\hbar \omega$$\simeq$2.4 meV and 4 meV. 
These excitations have the broad width 
as shown by the 1D-energy cut in Fig. \ref{Fig. 2a}(d).

In order to examine the $Q$-dependence of the excitation at 0.4 meV, 
the 1D-$Q$ cut obtained by the integration 
in the range of $0.3~{\rm meV} \le E \le 0.5~{\rm meV}$ 
is shown in Fig. \ref{Fig. 2a}(e). 
The dispersionless excitation 
has broad peak at $Q$=1.2 \AA$^{-1}$ of which the 
wave number corresponds to the distance between O$_2$ molecules. 
This excitation is not clearly observed in 0.3O$_{2}$-(Cu-CHD). 
The energy of the excitation is quite close to 
the anisotropy energy of the 
natural oxygen in the gas phase reported as 0.5 meV.\cite{Kimura} 
This suggested that the excitation is due to 
the single-ion-anisotropy of the adsorbed oxygen molecules. 

The 1D-$Q$ cut for the excitation at 4 meV 
is obtained by the integration range of 
$3.5~{\rm meV} \le E \le 4.5 {\rm meV}$ in Fig. \ref{Fig. 2a}(f).
The excitation has the peaks at about 1.2 \AA$^{-1}$ 
and the intensities decrease with increasing $Q$. 
The qualitative behavior is the same as that of 0.3O$_{2}$-(Cu-CHD), 
meaning that the origin of the excitation is 
a cluster of O$_2$ molecules.

The excitation at $\hbar \omega$$\simeq$2.4 meV 
exhibits stronger intensity 
at high $Q$. 
This means that its origin is nuclear lattice or cluster 
rather than magnetic one. 
We will exclude the excitation in the analysis section. 
Similarly we consider that 
the excitation at $\hbar \omega$$\simeq$2.4 meV in 0.3O$_{2}$-(Cu-CHD) 
would have nuclear origin.

\begin{figure}[t]
\begin{center}\leavevmode
\includegraphics[width=8 cm]{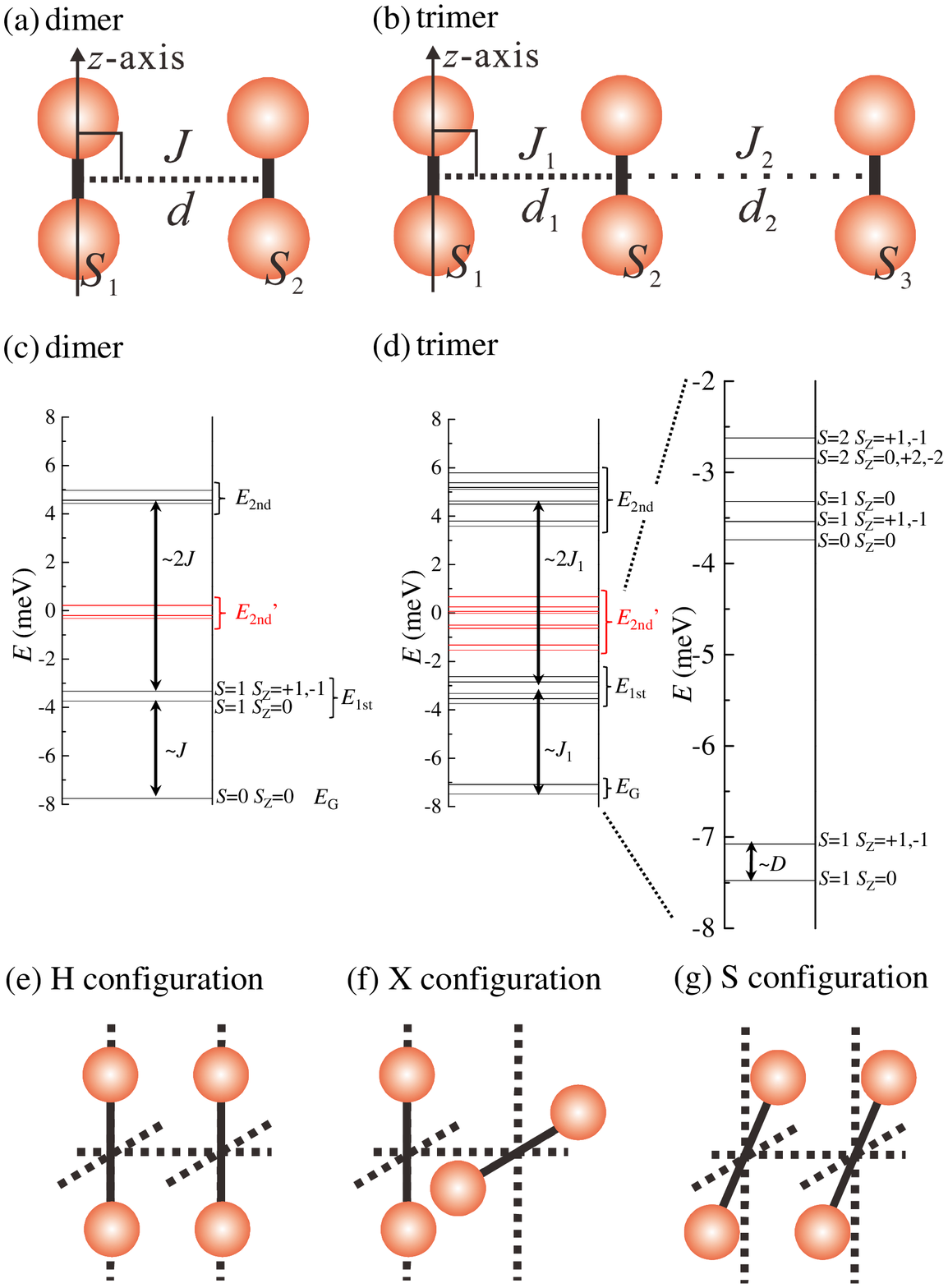}
\caption{
(Color online)
(a) O$_{2}$ configuration used in the calculation of the dimer model. 
(b) O$_{2}$ configuration used in the calculation of trimer model. 
(c) Black lines show energy levels calculated 
by conventional spin Hamiltonian of $S$=1 dimer model. 
Red lines show the reduced energy levels 
to reproduce the magnetization. 
(d) Black lines show energy levels calculated 
by conventional spin Hamiltonian of $S$=1 trimer model. 
Red lines show the reduced energy levels 
to reproduce the magnetization. 
(e) O$_{2}$ dimer in H-configuration. 
(f) O$_{2}$ dimer in X-configuration. 
(g) O$_{2}$ dimer in S-configuration.
}
\label{Fig. 3}
\end{center}
\end{figure}

\section{Analyses}
Combination of synchrotron x-ray diffraction and 
maximum entropy method/Rietveld analysis 
revealed that 
the adsorbed O$_2$ molecules form dimers in the nanopore.\cite{Hori} 
We, therefore, start our analysis on the magnetic excitation 
at $E$$\simeq$4 meV
in 0.3O$_{2}$-(Cu-CHD) from the $S=1$ spin-dimer model 
shown in Fig. \ref{Fig. 3}(a). 
We assume that the molecular axes of two O$_2$ molecules are parallel. 
The distance between the oxygen molecules, $d$, 
is fixed with the reported value 3.22 {\AA}.\cite{Hori} 
The magnetic anisotropy axis is along the molecular axis,~\cite{Tinkham} 
which is defined as $z$ - axis. 
The spin Hamiltonian is
\begin{eqnarray}
\mathcal{H}_{dimer}&=&J\mbox{\boldmath $S_{1}$}\cdot\mbox{\boldmath $S_{2}$}
+D\sum_{l}\left(S_{l}^{z}\right)^{2}.
\label{dimer}
\end{eqnarray}
The neutron cross section within the Born approximation~\cite{Squire} 
is calculated by the diagonalization of the Hamiltonian. 
The powder average was, then, performed to compare the calculation 
with the experiment. 
We used the magnetic form factor 
$F(\mbox{\boldmath $Q$})=
\int^{\infty}_{0}\int^{\pi}_{0}drd\theta r^{4}
\mbox{sin}^{3}\theta
 e^{-2br^{2}}[\mbox{cosh}(2bR_{0}\mbox{cos}\theta)-1]e^{iQr\mbox{cos}
(\theta-\beta)}$ as described in Ref. \onlinecite{formfactor}. 
Here, we used the internuclear distance $R$$_{0}$=1.21 {\AA} and  
a constant $b$=4.1 \AA$^{-2}$. 
$\beta$ is the angle between the molecular axis and the scattering vector.

Since the calculated neutron cross section of a spin cluster is 
a dispersionless delta function, the peak width of the excitation 
should be resolution limited. 
In contrast the excitation, which is guided by the 
green curve in Fig.~\ref{Fig. 1} (e), exhibits the larger width than 
that of the experimental resolution. 
The shape of the excitation is rather symmetric, and 
the asymmetry due to Van-Hove singularity is not observed. 
This suggests that the finite width is ascribed 
not to the powder-averaged dispersion due to 
an interdimer coupling 
but to the distribution of the intradimer 
exchange constants 
of isolated dimers. 
We consider that multiple minima for O$_2$ molecules 
are there inside the 
nanopore of Cu-CHD, leading to randomness of the 
configuration of O$_2$ dimer. 
Then the spin system is considered as 
a group of dimers having different $J$s. 
We assume that the distribution of the $J$s is Gaussian function. 
The free parameters for the fitting to the data 
are, thus, the mean value of the exchange interaction
redefined as $J$, 
and the standard deviation of the distribution $\sigma$. 
In this analysis, 
the uniaxial anisotropy $D$ is not very important 
because the magnitude of $\sigma$ is much larger than that of $D$. 
We fixed $D$ = 0.41 meV that is independently determined 
in the analysis of 2.0O$_{2}$-(Cu-CHD) which will be explained later. 
We also consider the additional background with the form of 
$A(\hbar \omega )Q^2 + B$. 


\begin{table}
 \caption{Parameteres obtained for the 0.3O$_{2}$-(Cu-CHD) sample from 
INS experiment (upper) and those for the 0.22O$_{2}$-(Cu-CHD) sample 
from magnetization measurement (lower).}
 \label{parameters}
\begin{center}
 \begin{tabular}{l l l l l l l}
 & $J$~(meV) &  $d$~(\AA) & $\sigma$~(meV) & $D$~(meV) & $N$(\%) 
& ${\Delta} E$\\
 \hline
INS & 4.15 & 3.22 & 2.16 & 0.41 & - & -\\
 \hline
$M$ Dimer & 4.15 & - & 1.77 & 0.41 & 90 & 1.15J \\
$M$ Monomer & - & - & - & 0.41 & 10 & - \\
 \hline
 \end{tabular}
 \end{center}
 \end{table}

For the fitting we use the data of one-dimensional energy cut 
in Fig.~\ref{Fig. 1a}(c) 
and one-dimensional $Q$ cut in Fig.~\ref{Fig. 1a}(d). 
In the first step we fit the one-dimensional energy cut 
by three Gaussians to obtain the peak energy and the 
energy width. 
We introduce the additional background with the phonon-like form 
i.e., 
$A_n{\exp}[-(({\hbar}{\omega}-{\hbar}{\omega}_n)/\sqrt{2}{\Delta}_n)^2]Q^2$ 
($n = 1, 2, 3$) and the additional flat background $B$. 
Here $\hbar {\omega}_n$ are the peak energies obtained in 
the first step. 
In the second step we fit 
the sum of the neutron cross section 
of the group of dimers and the additional background, 
$\sum _n (A_n{\exp}[-(({\hbar}{\omega}-{\hbar}{\omega}_n)
/\sqrt{2}{\Delta}_n)^2]Q^2) + B$ to 
the data of one-dimensional $Q$ cut
and one-dimensional energy cut. 
Fitting parameters are $J$, $\sigma$, scale factor for 
the cross section of group dimers, $A_n (n=1,2,3)$, and $B$. 

The calculated cross sections with the parameters 
summarized in Table I are shown by 
by the green solid curves in Figs. \ref{Fig. 1a}(c) 
and \ref{Fig. 1a}(d). 
The blue curves are the additional background, 
and the red solid curves 
are the sum of green and blue curves. 
The data are reasonably reproduced by the calculation, 
meaning that the magnetic excitation of the 0.3O$_{2}$-(Cu-CHD) 
at $T$ = 4.5 K is explained by the spin dimers 
with normally distributed exchange constants. 

The eigenenergies of a single dimer calculated by using 
the parameters of $J$ and 
$D$ in Table I are shown 
by the black lines in Fig. \ref{Fig. 3}(c). 
The ground state, $E_{G}$, is nonmagnetic singlet. 
The first group of the excited states, $E_{1st}$,
is composed of nearly triplet states with $S$=1, which are 
lifted by the single-ion anisostopy $D$. 
The second group, $E_{2nd}$, is composed of 
nearly quintet states with $S$=2. 
The transition between $E_{G}$ and $E_{1st}$ 
is probed as the excitation 
at $\hbar \omega = 4$ meV in the INS spectrum.

In the INS spectrum in 2.0O$_{2}$-(Cu-CHD) 
we observed a couple of magnetic excitations at 
$E$$\simeq$0.4 meV and 4 meV. 
The latter excitation looks similar to that observed in 
0.3O$_{2}$-(Cu-CHD), but the 
simple dimer model does not explain the 0.4 meV excitation. 
In fact the triplet state is the ground state 
in case that the cluster is composed of the odd number of 
Heisenberg spins, 
and, therefore, the transition between the triplet states lifted by 
the single-ion anisotropy $D$ can be probed as the low energy 
excitation. 
In the monomer case, the cross section 
decreases monotonically with increasing $Q$ 
because it depends solely on the magnetic form factor. 
In Fig.~\ref{Fig. 2} (g) the cross section exhibits a peak 
at $Q$$\simeq$1.2 \AA$^{-1}$, 
meaning that the cluster is composed of the multiple spins. 
We, thus, consider the trimer of O$_{2}$ molecules shown 
in Fig. \ref{Fig. 3}(b). 
We assume that the molecular axes are parallel. 
The trimer has two different inter-molecular lengths, 
$d_{1}$ and $d_{2}$, and correspondingly two different 
exchange constants, $J_1$ and $J_2$.
The spin Hamiltonian of a trimer is, thus, 
\begin{eqnarray}
\mathcal{H}_{trimer}&=&J_{1}\mbox{\boldmath $S_{1}$}
\cdot\mbox{\boldmath $S_{2}$}
+J_{2}\mbox{\boldmath $S_{2}$}\cdot\mbox{\boldmath $S_{3}$}
+D\sum_{l}\left(S_{l}^{z}\right)^{2}.
\label{trimer}
\end{eqnarray}
The neutron cross section is calculated by the diagonalization of 
the Hamiltonian. 
The free parameters are the exchange interactions $J_1$, $J_2$, 
the uniaxial anisotropy $D$, and one of the inter-molecular distances 
$d_2$. 
Another distance, $d_{1}$, is fixed to 3.22 {\AA}. \cite{Hori} 
We assume that the width of the magnetic excitation at $E$$\simeq$4 meV 
is due to the normal distribution of the exchange constant $J_{1}$. 
We define that the standard deviation $\sigma_{1}$ of the 
distribution, and redefine the mean value of the 
exchange constants as $J_1$. 
Both of them are the free parameters for the calculation. 
The standard deviation $\sigma_{2}$ of $J_{2}$ is fixed to 0. 
The additional background 
is also considered in a similar way to 
the analysis in 0.3O$_2$-(Cu-CHD). 

The calculated cross sections 
are shown by the solid curves 
in Figs. \ref{Fig. 2}(e)-\ref{Fig. 2}(h), 
and the obtained parameters are shown in Table II. 
The green curve is the magnetic excitation, 
the blue curve is additional background, 
and the red curve is the sum of green and blue curves. 
The data is reasonably reproduced by the spin trimers of which 
the main exchange constants are normally distributed. 
We note that the ratio of the peak intensity 
at $Q$$\simeq$1.2 \AA$^{-1}$ 
to the amplitude of the intensity modulation 
in Fig.~\ref{Fig. 2} (g) is tuned by the ratio 
of $J_{1}$ to $J_{2}$. 
The value of $d_{2}$ is larger than that of $d_{1}$, 
and the value of $J_{2}$ is much smaller than that of $J_{1}$. 
This means that the O$_2$ trimers 
 in the high oxygen concentration sample 
2.0O$_{2}$-(Cu-CHD) are regarded as 
the O$_{2}$ dimers weakly coupled to 
O$_2$ monomers. 

The energy level of a single trimer 
calculated by the obtained parameters 
and the mean exchange constant $J_1$ 
are shown by the black lines in Fig. \ref{Fig. 3}(d). 
The eigenstates of the trimer model are separated 
into the three groups. 
The group having the low energy, $E_{G}$, 
is $S$ = 1 states lifted by the single-ion anisotropy $D$. 
They are composed of 3 states and the ground state is $S_{z}$=0. 
The first group of the excited states, $E_{1st}$,
is composed of 9 states, 
and the second group, $E_{2nd}$, 
is composed of 15 states.
The excitation at $E$$\simeq$0.4 meV corresponds to 
the transitions among the $E_{G}$ states. 
The excitation at $E$$\simeq$4 meV 
corresponds to the transitions 
between $E_G$ states and $E_{1st}$ states. 

The $Q$ dependence of the excitation at 4 meV in 2.0O$_2$-(Cu-CHD) 
is similar to that in 0.3O$_2$-(Cu-CHD), since they are physically 
quite similar. 
If $J_2$ is zero and the system is 
composed of isolated dimers and monomers, 
the transition between $E_{G}$ and $E_{1st}$ in the system 
is exactly the same as that in dimers system. 
It is, therefore, rather difficult to exclude the possibility 
that some amount of dimers are included in 2.0O$_2$-(Cu-CHD). 
Even though the intensity comparison of the excitation at 0.4 meV, 
which is purely from spin timer, and 
that at 4 meV tells us the ratio of dimers to trimers in principle, 
the separated data set in the different experimental setups makes 
the comparison difficult.

\begin{table*}
 \caption{Parameteres obtained for the 2.0O$_{2}$-(Cu-CHD) sample from 
INS experiment (upper) and those for the 1.11O$_{2}$-(Cu-CHD) sample 
from magnetization measurement (lower).
}
 \label{parameters}
\begin{center}
 \begin{tabular}{l l l l l l l l l l}
& $J_{1}$~(meV) &  $d_{1}$~(\AA) & $\sigma_{1}$~(meV)
& $J_{2}$~(meV) &  $d_{2}$~(\AA) & $\sigma_{2}$~(meV) & $D$~(meV) 
& $N$(\%) & ${\Delta}E$\\
 \hline
INS & 3.94 & 3.22 & 1.54 & 0.62 & 3.5 & 0 & 0.41 & - & - \\
 \hline
$M$ Trimer & 3.94 & - & 2.52 & 0.62 & - & 0 & 0.41 & 33 & 1.3$J$ \\
$M$ Dimer & 4.15 & - & 3.54 & - & - & - & 0.41 & 67 & 1.3$J$ \\
 \hline
 \end{tabular}
 \end{center}
 \end{table*}

\section{Discussion}

\begin{figure*}
\begin{center}\leavevmode
\includegraphics[width=15 cm]{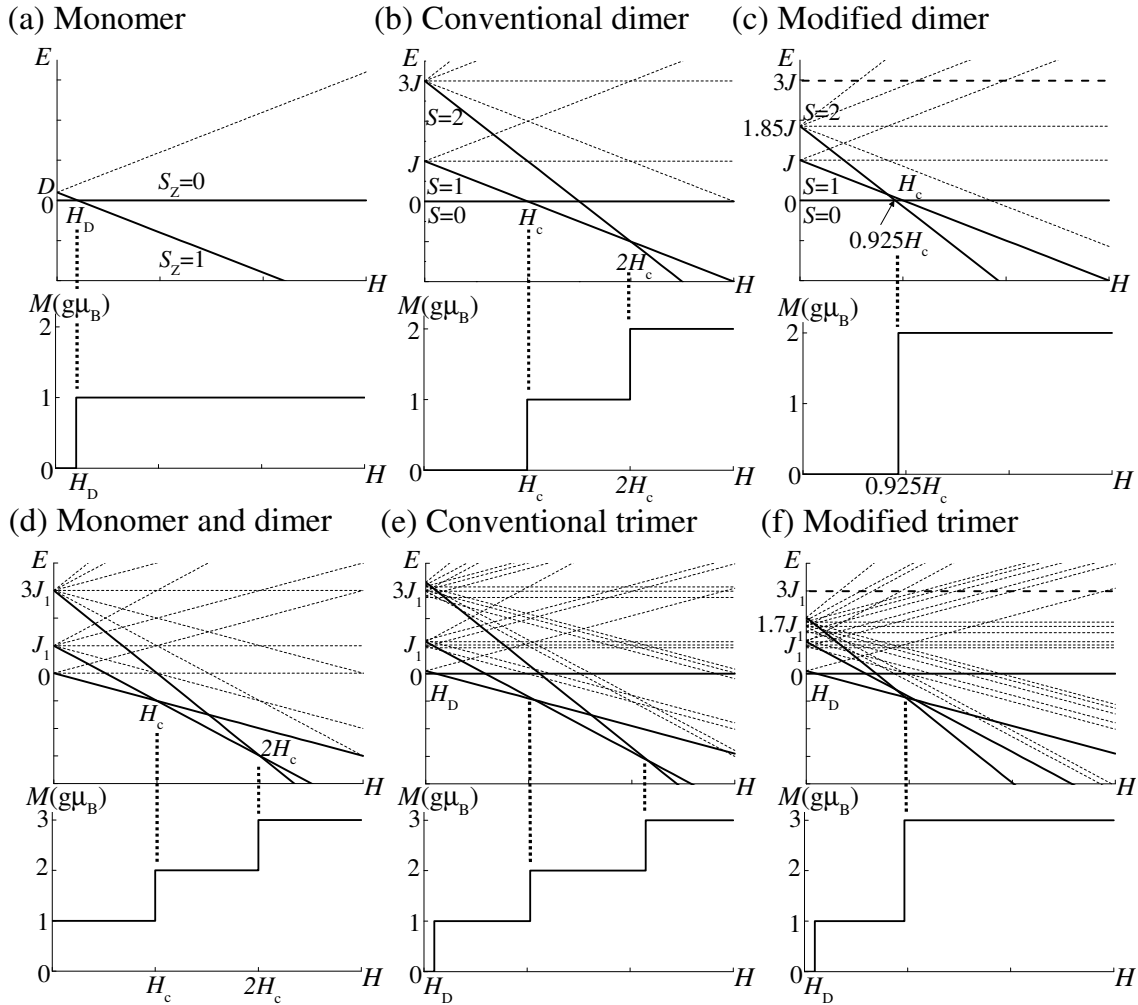}
\caption{
The field dependence of the energy levels and corresponding 
magnetization curve in $S$ = 1 spin cluster systems. 
(a) Monomer spin with a single-ion anisotropy D. 
(b) Conventional Heisenberg spin dimer. 
(c) Heisenberg spin dimer with the reduced energy levels of $S$ = 2 
states. 
(d) Isolated monomer and dimer. 
(e) Conventional trimer where a dimer is weakly coupled to a monomer. 
Single-ion anisotropy is also considered. 
(f) The trimer with the reduced energy levels of
higher energy states. 
}
\label{Fig. 4}
\end{center}
\end{figure*}

\begin{figure}
\begin{center}\leavevmode
\includegraphics[width=8 cm]{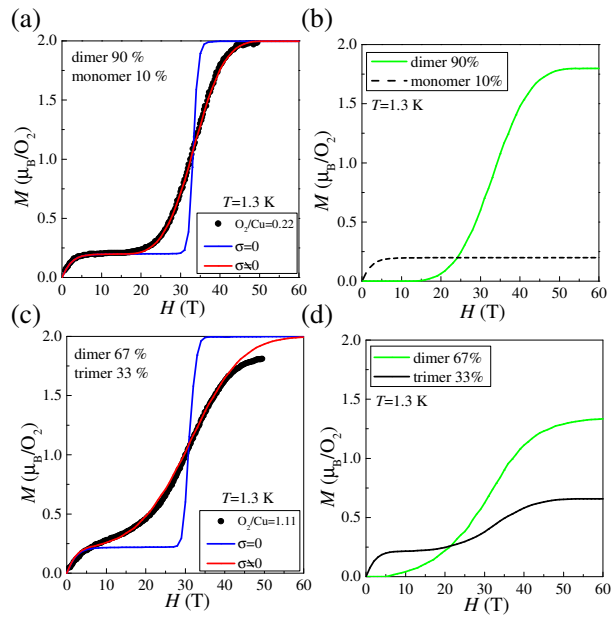}
\caption{
(Color online) Comparison between the magnetization 
reported by W. Mori $et$ $al.$\cite{Mori} 
and the calculated magnetization. 
The black symbols show the reported data, 
and the blue and red solid lines represent the calculation (see text). 
(a) Magnetic field dependence of the magnetization 
for Cu-CHD adsorbing less oxygen molecules. 
(b) Temperature dependence of the magnetic susceptibility 
for Cu-CHD adsorbing less oxygen molecules. 
(c) Magnetic field dependence of the magnetization 
for Cu-CHD adsorbing more oxygen molecules. 
(d) Temperature dependence of the magnetic susceptibility 
for Cu-CHD adsorbing more oxygen molecules.
}
\label{Fig. 5}
\end{center}
\end{figure}

The magnetic field dependence of the energy level of 
Heisenberg $S$=1 spin dimer is shown in the upper panel of 
Fig.~\ref{Fig. 4} (b), and the corresponding 
magnetization curve is shown in the lower panel. 
The magnetization curve exhibits $1/2$ plateau at 
$H_c \le H \le 2H_c$ with $H_c$ = $J/(g{\mu}_B),$ 
where the ground state is 
$\lvert S, S^z \rangle = \lvert 1, 1 \rangle$. 
In contrast in 0.22O$_{2}$-(Cu-CHD) the plateau 
is absent\cite{Mori} as shown in Fig.~\ref{Fig. 5}(a), 
and it seems more likely two-level system. 
Anomalous energy level of $S=1$ spin dimers in the adsorbed oxygen system 
was initially discussed in the temperature dependence of 
the neutron intensity in a 
different metal complex, CPL-1.\cite{Masuda}
The intensity decreased more rapidly 
with the increase of the temperature 
compared with conventional $S$ = 1 spin dimer model, 
suggesting the lowering of the energy level of the $S$ = 2 
quintet state and the proximity to the triplet state of $S$ = 1. 
The energy level was explained based on the calculation including
 inter-molecular potential of O$_2$.\cite{Bussery} 
For the dimer of O$_{2}$ molecules, 
both the ground state with $S$=0
and the first excited states with $S$=1
have an H-configuration shown in Fig. \ref{Fig. 3}(e) 
in the equilibrium molecular positions.
 On the other hand, the second excited state with $S$ = 2 
has an X-configuration shown in Fig. \ref{Fig. 3}(f)
of which the energy is lower than that in 
the H-configuration. 
The energy difference between $S$ = 0 and $S$ = 1 in both H configuration 
and that between $S$ = 1 in H configuration and 
$S$ = 2 in X configuration is almost the same, 
resulting in the anomalous energy level. 

In the magnetization and x-ray study in 0.22O$_{2}$-(Cu-CHD), 
the lowering of the $S$ = 2 state was explained in the similar 
manner.\cite{Hori} 
According to their Rietveld/MEM analysis $S$-geometry 
as shown in Fig. \ref{Fig. 3}(g) 
was realized for $S$ = 2 state. 
It was estimated that the first energy-gap between $S=0$ and $S=1$ 
is 3.7 meV and the second one between $S=1$ and $S=2$ was 
4.7 meV. 
The distribution of the exchange constant and the single-ion anisotropy 
term were not considered there. 
In the present paper we consider the magnetization data of both 
O$_{2}$/(Cu atom)=0.22 and 1.11 reported by W. Mori $et$ $al.$ \cite{Mori}
Our discussion is based 
on the spin monomer/dimer/trimer model with the normally distributed 
exchange constant and the 
single-ion anisotropy obtained from the INS spectra. 
We also assume that the eigenenergies of higher states 
calculated by conventional spin Hamiltonian 
are reduced because of the spin-dependent molecular potential 
in the oxygen dimer.\cite{Hemert,Wormer,Bussery,Bussery94} 

We consider the following Hamiltonians 
\begin{eqnarray}
\mathcal{H}_{\rm mono}&=&D\sum_{l}\left(S_{l}^{z}\right)^{2}
-g\mu_{B}H\sum_{l}S_{l}^{z'}\\
\mathcal{H}_{\rm dimer}&=&J\mbox{\boldmath $S_{1}$}\cdot\mbox{\boldmath $S_{2}$}
+D\sum_{l}\left(S_{l}^{z}\right)^{2}-g\mu_{B}H\sum_{l}S_{l}^{z'}\\
\mathcal{H}_{\rm trimer}&=&J_{1}\mbox{\boldmath $S_{1}$}
\cdot\mbox{\boldmath $S_{2}$}
+J_{2}\mbox{\boldmath $S_{2}$}
\cdot\mbox{\boldmath $S_{3}$}
+D\sum_{l}\left(S_{l}^{z}\right)^{2}  \nonumber \\
&-&g\mu_{B}H\sum_{l}S_{l}^{z'}
\end{eqnarray}
where the direction of the magnetic field, $H$, is defined as $z'$-axis. 
For all models, we fixed $D$=0.41 meV. 
The magnetizations were calculated using the above Hamiltonians 
having the distributions in $J$ and $J_1$. 
The powder average was performed.

The reported magnetization in 0.22O$_{2}$-(Cu-CHD) exhibited 
a small increase below 5 T and a large increase at around 20 T 
with increasing $H$. 
The former increase is not explained by $S$ = 1 dimer model but 
by $S$ = 1 monomer model. 
As schematized in Fig. \ref{Fig. 4}(a) the magnetization 
of the monomer at $T$ = 0 K 
exhibits a 
step function of which the jump field is $H_D = D/(g{\mu}_B)$. 
Use the value of $D$ = 0.41 meV and $H_D$ = 3.5 T is obtained. 
At finite temperature the jump is smeared. 
The field scale of the initial increase of the magnetization is, 
thus, consistent with that of 
the single-ion anisotropy $D$ = 0.41 meV. 
We, therefore, assume that $S$ = 1 oxygen molecule 
monomers were trapped in some 
local minima in the nanopore or surface of the grain of the 
crystal in 0.22O$_{2}$-(Cu-CHD). 
The ratio of O$_2$ molecules in the monomer, $N$, 
is estimated to be 10 \% of the total O$_2$ molecules from the 
magnetization plateau at $8 {\rm T} \lesssim H \lesssim 15 {\rm T}.$ 
The small number of the monomers would be the reason why the excitation 
was not observed in INS spectrum. 

The red lines in Fig. \ref{Fig. 3}(c) 
shows the energy levels of the spin dimer of which the energy 
of the quintet $S$ = 2 state is lowered compared with 
the conventional spin dimer. 
The field dependences of the conventional and modified 
states are shown in 
the upper panel in Fig. \ref{Fig. 4}(b) and (c). 
The single-ion anisotropy is omitted 
for the simplicity here. 
In the modified scheme 
the energy of the $S^z$ = 2 crosses at the lower field than 
that of $S^z$ = 1 does, meaning that the magnetization 
curve is a single-step function 
as shown in the lower panel. 
By introducing the distribution of the exchange constant, 
the critical field is smeared and the experimentally obtained 
magnetization-curve 
is reproduced. 
We assume that the inflection point of the experimental magnetization 
curve is the critical field of the spin dimer having the mean value of 
$J$, 4.15 meV, in Table I. 
The field of the inflection is estimated to be 34 T, leading to that 
the decrease of the energy of the $S$ = 2 state, ${\Delta}E$, 
is 1.1$J$. 
Considering the single-ion anisotropy we found that 1.15$J$ is 
more appropriate value. 
The distribution of the $J$ is manually changed to find the best
consistency. 
It is find that 1.77 meV is the best standard deviation of $J$. 
The magnetization curve calculated by using the parameters 
in Table I is indicated by the red solid curve in Fig. \ref{Fig. 5}(a). 
The data is reasonably reproduced by the calculation. 
The blue solid curve is the magnetization for $\sigma$ = 0, meaning that 
the distribution of $J$ is required. 
The component of dimer and that of monomer is indicated by 
green solid curve and black dashed curve in Fig. \ref{Fig. 5}(b). 

The lowering of the $S$ = 2 state is not probed in the INS profile 
because the transition between $S$ = 0 and $S$ = 2 is forbidden. 
The transition between $S$ = 1 and $S$ = 2 is not forbidden but is 
not observed because of the small probability of $S$ = 1 state at 
the low temperature. 
We note that the excitation at $E$$\simeq$2.4 meV observed in INS 
is not probed in the magnetization. 
The result supports our consideration 
that it is not from the magnetic origin.

The magnetization of 1.11O$_2$-(Cu-CHD) measured by 
Mori {\it et al.}~\cite{Mori} is indicated by solid symbols 
in Fig. \ref{Fig. 5}(c). 
To understand the qualitative behavior of the $S$ = 1 trimer composed 
of a dimer weakly coupled to a monomer, 
it is instructive to consider an isolated monomer and dimer. 
The energy level and the corresponding magnetization curve 
is shown in Fig. \ref{Fig. 4}(d). 
The magnetization exhibits three-step staircase where the critical fields are
$H_{1,c}$ and $2H_{1,c}$ with $H_{1,c} = J_1/(g{\mu}_B)$. 
The ground state is $S$ = 1 triplet and the magnetization has 
1 $g{\mu}_B$/trimer even at 0 T. 
By introducing the weak coupling between dimer and monomer $J_2$ and 
the single-ion anisotropy $D$, the degenerated states are lifted and 
the energy levels and the corresponding magnetization curve are modified 
as shown in Fig. \ref{Fig. 5}(e). 
Here the highest states are $\lvert S, S_z \rangle =\lvert 3, \pm 3 \rangle$ 
and the energy is about $3J_1$. 
In order to make the magnetization a two-step staircase, 
it is necessary to lower the energies of 
the group of the higher states at about $3J_1$ 
by ${\Delta}E$ as shown in the upper panel in Fig. \ref{Fig. 4}(f). 
By introducing the distribution of $J_1$ the step function is smeared. 
We tried to reproduce the experimental data 
by the trimers model but we failed. 
Instead the data was reproduced by the sum of the 
distributed dimers and the distributed trimers 
as indicated by red solid curve in Fig. \ref{Fig. 5}(c). 
The ratio of O$_2$ in dimers is 67\% of the total adsorbed 
O$_2$. 
In Fig. \ref{Fig. 5}(d) 
the magnetization of the dimers and trimers are indicated by 
green and black solid curves, respectively. 
The parameters used in the calculation are summarized in Table II. 

The concentration of O$_2$ in 
the sample used in magnetization measurement is 1.11[mole/f.u.], 
and it is smaller than that in INS measurement 2.0[mole/f.u.]. 
The less amount of O$_2$ would be the reason for the 
substantial number of dimers in the magnetization data. 
The data, however, indicates that 
some amount of dimers may be included also in 
the sample of 2.0O$_2$-(Cu-CHD) 
used in the INS experiment. 
The existence of several potential minima 
in the nanopore may be a possible reason for 
the inclusion of both dimers and trimers. 

\section{Conclusion}
We performed INS measurements 
on Cu-CHD adsorbing O$_{2}$ molecules 
with the low and high concentrations 
to identify the magnetism of the O$_{2}$-based magnet
realized in the nanopores. 
The INS spectra of the two different samples are 
explained by different spin Hamiltonians, 
spin dimers and spin trimers, meaning that the 
spin system can be controlled by the concentration of O$_2$ molecule 
in the O$_{2}$-based magnet. 
It is found that the magnetic excitation is broad 
and, therefore, the system is composed of a group of clusters having 
normally distributed exchange constants. 
This indicates that the supercrystal of the oxygen molecules is not 
perfect. 
By using the parameters obtained in INS and by assuming the reduction 
of the higher energy states due to the 
non-negligible spin-dependence in the molecular potential, 
magnetization curves are explained in quantitative level. 

\section*{Acknowledgment}

The neutron scattering experiment was approved 
by the Neutron Science Proposal Review Committee of J-PARC/MLF (2012B0067). 
The work is supported by Grants-in-Aid for Scientific Research 
from the Japan Society for the Promotion of Science (JSPS) 
and by Grants-in-Aid on priority areas 
from the Ministry of Education, Culture, Sports, Science and Technology.
 This work was partially supported 
by Grants-in-Aid for Scientific Research KAKENHI (24340077).

\clearpage

\end{document}